\documentclass[10pt,twocolumn,preprintnumbers,amsmath,amssymb,prl,aps,floatfix,superscriptaddress,showpacs]{revtex4-1}

\usepackage{epsfig}
\usepackage{graphicx}
\usepackage{graphicx}
\usepackage{dcolumn}
\usepackage{bm}
\usepackage{hyperref}
\begin{document}

\title{Low Temperature Studies of Charge Dynamics of Nitrogen-Vacancy Defect in Diamond}

\author{P. Siyushev}
\email[]{p.siyushev@physik.uni-stuttgart.de}
\affiliation{3.Physikalisches Institut and Stuttgart Research Center of Photonic Engineering (SCoPE), Universit\"at Stuttgart, Pfaffenwaldring 57, Stuttgart, D-70569, Germany}
\author{H.Pinto}
\affiliation{Institute for Solid State Physics and Optics, Wigner Research Center for Physics, Hungarian Academy of Sciences, Budapest, POB 49, H-1525, Hungary}
\author{A.Gali}
\affiliation{Institute for Solid State Physics and Optics, Wigner Research Center for Physics, Hungarian Academy of Sciences, Budapest, POB 49, H-1525, Hungary}
\affiliation{Department of Atomic Physics, Budapest University of Technology and Economics, Budafoki ut 8, H-1111, Budapest, Hungary}
\author{F. Jelezko}
\affiliation{Institut f\"ur Quantenoptik, Universit\"at Ulm, D-89081 Ulm, Germany}
\author{J. Wrachtrup}
\affiliation{3.Physikalisches Institut and Stuttgart Research Center of Photonic Engineering (SCoPE), Universit\"at Stuttgart, Pfaffenwaldring 57, Stuttgart, D-70569, Germany}
\affiliation{Max Planck Institute for Solid State Research, Heisenbergstra\ss{}e 1, D-70569 Stuttgart, Germany}

\begin{abstract}

 In this paper, we study the photoinduced switching of the nitrogen-vacancy (NV) center between two different charge states -- negative (NV$^{-}$) and neutral (NV$^{0}$) at liquid helium temperature. The conversion of NV$^{-}$ to NV$^{0}$ on a single defect is experimentally proven and its rate scales quadratically with power under resonant excitation. In addition, we found that resonant excitation of the neutral NV changes the charge state, recovering its negative configuration. This type of  conversion significantly improves spectral stability of NV$^{-}$ defect and allows high fidelity initialization of the spin qubit. A possible mechanism for ionization and recovery of the NV$^{-}$ defect is discussed. This study provides better understanding of the charge dynamics of the NV center, which is relevant for quantum information processing based on NV defect in diamond.

\end{abstract}


\maketitle
\begin{figure}[]
\centerline
{\includegraphics[width=0.5\textwidth]{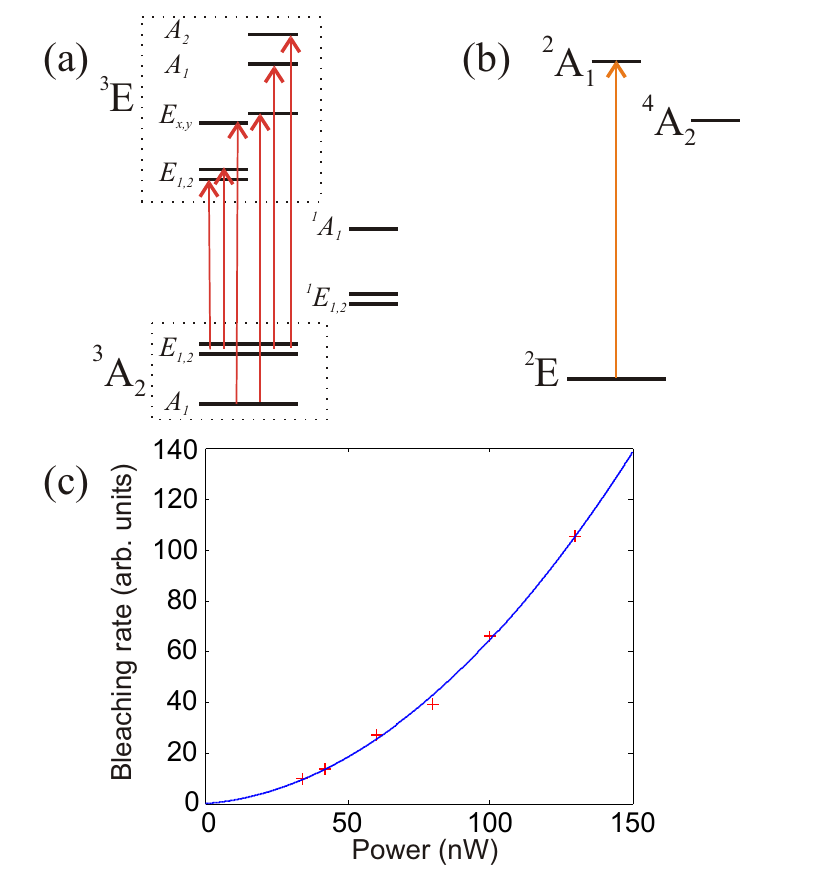}}
\caption{\label{fig1}Energy-level diagram of the NV defect in the (a) negative and (b) neutral charge state. (c) Bleaching rate of the NV$^-$ as a function of resonant laser power. Each point is a number of switching events extracted from six scans at certain laser power of the red laser. Single scan consists of one thousand laser sweeps across the same NV$^{-}$ transitions.}
\end{figure}

\indent The negatively charged nitrogen-vacancy (NV$^{-}$) center is a leading contender~\cite{Ladd_2010} in solid-state quantum information processing due to long spin coherence time (up to several milliseconds in ultra-pure diamonds~\cite{Balasubramanian_2009}) and ease of optical initialization and readout~\cite{Jelezko_2006}, even non-destructively~\cite{Awschalom_Nature2010,Robledo_Nature2011}. In particular, it is of potential use for quantum repeaters because it combines strong optical transitions with long lived spin states. Such applications based on long-distance entanglement protocols require indistinguishable photons. It has been shown that the NV defect emits Fourier-transform limited photons~\cite{Batalov_PRL2008}, however, the main test for indistinguishability is Hong-Ou-Mandel (HOM) interference, recently demonstrated with two remote NV centers~\cite{Sipahigil_arxiv,Bernien_arxiv}. The contrast of two-photon interference is significantly deteriorated indicating that the probability to generate a pair of indistinguishable photons is low. One of the factors that reduces the contrast of HOM interference is spectral instability. In addition, the defect exists in various charge states, which are determined by surrounding impurities~\cite{Rondin_PRB2010,Fu_APL2010} and the position of the Fermi level~\cite{Hauf_PRB2010}, but are also influenced by the incident laser field at usual operation~\cite{Gaebel_2006,Waldherr_PRL2011}. Therefore, emission from undesired charge states decreases the number of indistinguishable photon pairs. In spite of the fundamental importance of understanding the mechanism of charge state switching, there is still no comprehensive picture of this process.\\
\indent In this letter, we report the charge dynamics of single NV defects at cryogenic conditions. We show that the switching rate of the NV$^{-}$ defect into a non-fluorescent state under resonant excitation has a quadratic power dependence. By detecting the photoluminescence excitation (PLE) spectrum of neutral nitrogen-vacancy (NV$^{0}$) defect, we prove that photoionization is the process responsible for turning NV into a dark state. We demonstrate a new approach for NV$^{-}$ fluorescence recovery, which helps significantly improve the spectral stability of the defect in the negatively charged state. Moreover, the NV defect can be deterministically prepared in a desirable charge state. A model for the photoionization and recovery mechanism is presented, which is supported by hybrid density functional theory calculations.\\
\indent The NV defect consists of a substitutional nitrogen atom and a vacancy at an adjacent lattice site. The {\it C$_{3v}$} point group symmetry imposed by the diamond lattice together with the number of electrons defines the defect's properties. In the negative charged state, the NV defect contains six electrons which, according to defect-molecule modeling are distributed in the $a_1^{(2)}$$a_1^{(2)}$$e^{(2)}$ electronic configuration~\cite{Lenef_PRB1996}. From these orbitals, triplet and singlet many-body states are formed. The corresponding levels are shown in Fig.~\ref{fig1} (a). The ground state is represented by the spin triplet $^{3}$A$_{2}$. Due to spin-spin interaction ground state spin sublevels are split by 2.87~GHz. The excited state $^{3}$E is an orbital doublet with degeneracy lifted by nonaxial strain~\cite{Batalov_PRL2009,Tamarat_NJP2008}. Each orbital branch is a spin triplet, whose sublevels are optically coupled to the ground state by spin preserving selection rules, therefore under resonant excitation the NV$^{-}$ defect should reveal six lines. However, only two sublevels {\it E$_{x}$} and {\it E$_{y}$}, coupled to the {\it A$_{1}$} in the ground state, involve good cycling transitions. Four other sublevels of the $^{3}$E state can decay to the metastable state due to spin-orbit interaction followed by decay to {\it A$_{1}$} sublevel of the ground state. This mechanism underlies in optical spin initialization process.\\
\indent In the case of the neutral charge state, NV possesses only five electrons, forming the level structure depicted in Fig.~\ref{fig1} (b). In contrast to the NV$^{-}$ defect, NV$^{0}$ is poorly investigated, but theoretical analysis and EPR studies~\cite{Felton_PRB2008,Gali_PRB2009} suggest that the $^{2}$E ground state is optically linked to the $^{2}$A$_{1}$ excited state, from which shelving is possible to the lower lying $^{4}$A$_{2}$ excited state.\\
\indent To perform the experiments, an ultrapure synthetic IIa diamond, grown by chemical vapor deposition (CVD) technique with nitrogen concentration below 5~ppb was used. Naturally occurring NV centers are well spatially separated and exhibit good spectral stability. Individual color centers were addressed by conventional confocal microscopy at low temperature ($T\approx 2$ K). A tunable red laser (extended cavity diode laser), emitting at $\approx 637$~nm, was used to excite the NV$^{-}$ defect resonantly at its zero phonon line (ZPL). PLE scans around the ZPL of NV$^{0}$, 575~nm, were carried out with a yellow laser (a tunable rhodamine~6G dye ring laser) operating in single mode. Excitation light was focused onto the sample by a microscope objective lens with NA=0.85, immersed in liquid helium. The same lens collects fluorescence which is filtered from the reflected pump light by a bandpass filter with transmission in the range of 650 - 750~nm. Also, a microwave field at 2.87~GHz was continuously applied via a 20~$\mu$m thick copper wire spanned across the diamond, in order to mix the population of the NV$^{-}$ ground state sublevels and thus maintain fluorescence of the non-cycling transitions.\\
\indent In order to describe the charge dynamics of defects under optical
excitation, density functional theory calculations were performed
using HSE06 screened Hartree-Fock hybrid for the exchange-correlation
functional as implemented in VASP
code~\cite{Kresse_PRB1996,Paier_JChP2006}. The atomic cores were
replaced by PAW-type
potentials~\cite{Bloechl_PRB1994,Joubert_PRB1999} while the valence
electrons were represented by a plane waves basis set with a cutoff
energy of 420~eV. The Brillouin zone was sampled using the $\Gamma$
point ensuring well-converged charge density.\\
\indent For the study of the charge dynamics a single NV defect was
placed in a 512-atom supercell. The structure for each electronic
configuration was optimized until the forces acting on each atom were
less than 0.01~eV/{\AA}. The optical transitions were obtained from
differences in total energies between the different electronic
configurations.  Within this method and with the used Brillouin
sampling and supercell size, the valence and conduction edges are well
described. The calculated gap between the highest occupied and lowest
unoccupied Kohn-Sham eigenvalues is 5.38~eV which can be compared with
the experimentally observed value of 5.48~eV. This method has previously been
used to successfully reproduce excitations processes of the NV$^{-}$
defect in diamond~\cite{Gali_PRL1983}. The single Kohn-Sham states
were also analysed through the projected density of states and the
transition dipole moment between
them was calculated.\\
\indent As previously reported~\cite{Fu_APL2010}, resonant excitation can easily suppress emission of the negatively charged NV defect, transferring it to a long living dark state. To restore fluorescence, shorter wavelength, deshelving laser pulses need to be applied. With increasing resonant laser power, the number of switching events into a non-fluorescing state rises. This rate is measured as a function of resonant laser power and the number of switching events is plotted in Fig.~\ref{fig1}(c). Experimental data are well fitted by a power function with exponent $1.9\pm0.1$, which is very close to the quadratic dependence. This suggests that process involves two photons. A similar behaviour was recently reported in Ref.~\cite{Waldherr_PRL2011} at room temperature.\\
\begin{figure}[]
\centerline
{\includegraphics[width=0.5\textwidth]{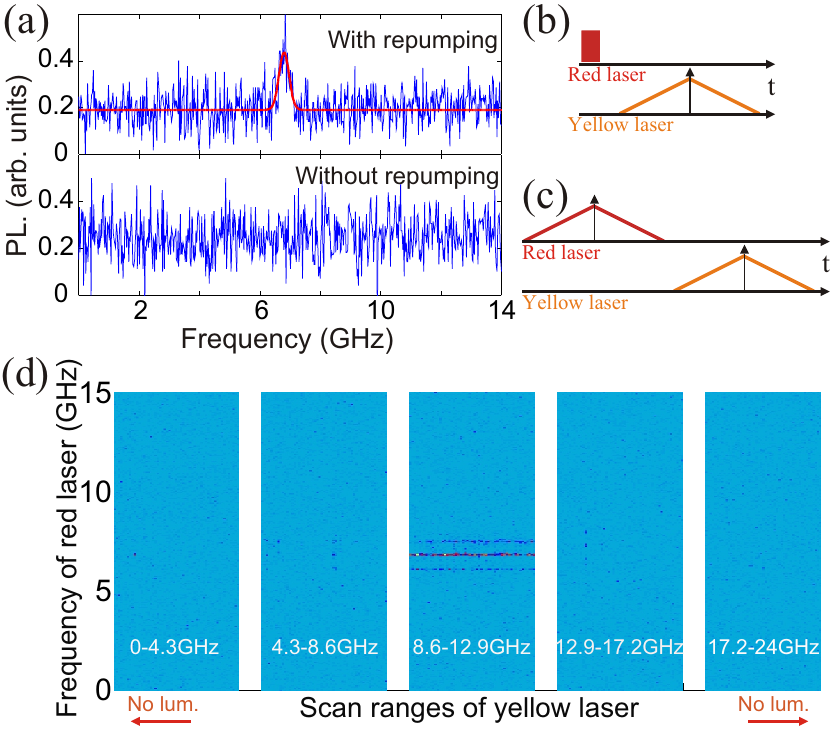}}
\caption{\label{fig2}(a) PLE spectrum of NV$^{0}$ defect with ZPL centered around 575.015~nm. The spectrum is observed only when photoionizing laser pulses are apllied (upper plot). (b) Sequence of ionizing red laser pulse at the resonance with one of  NV$^{-}$ transitions followed by frequency sweep around ZPL of NV$^{0}$. (c) Sequence used for the conversion NV$^{0}$ into NV$^{-}$. (d) Effect of fluorescence recovery on NV$^{-}$ transitions (only upper branch is shown), while additional laser is sweeping over the ZPL of NV$^{0}$. Each bar corresponds to one particular sweep range of the yellow laser. The scan for two nanometers (574 - 576~nm) is conducted with laser power of 100~nW and repumping is observed only for the narrow region (middle bar) and it does not appear for all other intervals.}
\end{figure}
\indent Despite many assumptions and some experimental investigation~\cite{Manson_2005}, there is no experimental proof on single NV centers, that resonant bleaching of NV$^{-}$ emission results in its charge conversion. If this process takes place, it should be possible to detect the neutral charge state. For that purpose, the single-frequency yellow laser was swept over the NV$^{0}$ resonance. To make sure, that NV$^{-}$ is neutralized with high probability, high power red laser pulses, tuned in resonance with one of the NV$^{-}$ transitions (for instance {\it A$_{1}$} $\rightarrow$ {\it E$_{x}$}), were applied at the beginning of each sweep (Fig.~\ref{fig2}(b)). A faint, broad PLE line at 575.015 nm is seen, which disappears as soon as the frequency of the ionizing red laser is shifted away from the NV$^{-}$ transition (see Fig~\ref{fig2}(a)). This experiment confirms that resonant excitation neutralizes the NV$^{-}$ defect. The ZPL of the NV$^{0}$ center exhibited strong spectral instability leading to a linewidth of a few hundred megahertz, slightly varying from one center to another.\\
\begin{figure}[]
\centerline
{\includegraphics[width=0.5\textwidth]{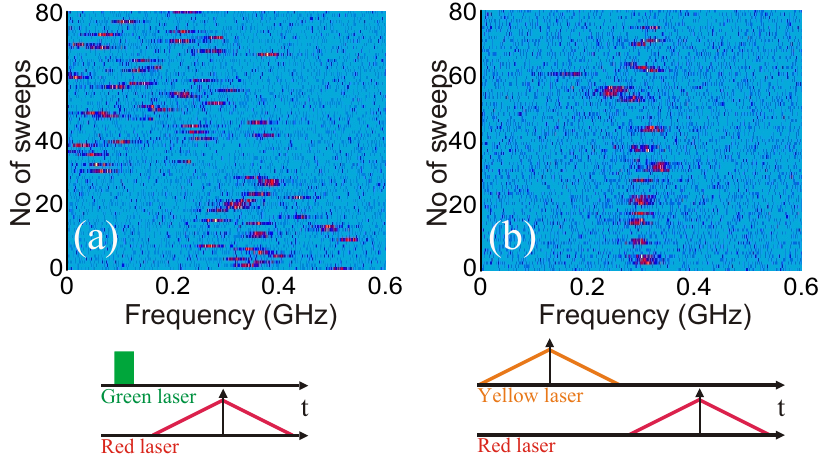}}
\caption{\label{fig3} PLE spectra of NV$^{-}$ with two different types of repumping: (a) conventional way, when at the beginning of each laser frequency sweep over NV$^{-}$ transitions a 1~mW green laser pulse is applied and (b) resonant repumping, when one laser is swept across ZPL of the NV$^{0}$ transition with power of 100~nW, before the red laser sweep.}
\end{figure}
\indent A surprising effect was found when an
alternative method for NV$^{-}$ neutralization was used, consisting of two consecutive
sweeps. First, a sweep of the red laser over all NV$^{-}$
resonant lines with an optical power of $\sim$10~$\mu$W was used in order to photoionize
the center. Second, the frequency of the yellow laser was swept across the NV$^{0}$ ZPL (Fig.~\ref{fig2}(c)). Stokes-shifted fluorescence was monitored
synchronously for both laser sweeps. As in the previous
experiment, a PLE spectrum of NV$^{0}$ appeared, but in addition, fluorescence of
NV$^{-}$ was observed upon the red laser excitation. Scanning of the yellow laser
through the ZPL of NV$^{0}$ causes a strong `repumping' effect back to
its negative charge state even at very low power (5~nW). Contrary to
the photoionization process of NV$^{-}$, where an electron is promoted
into the conduction band, NV$^{0}$ acquires an
electron from the lattice. The current concept~\cite{Kok_2006,Hanson_2010} is that the additional electron trapped by
the NV defect is acquired from substitutional nitrogen defects with
excitation threshold above 1.95~eV (637~nm). This means that once
a certain activation energy is exceeded, nitrogen should always be ionized.
This is not the case in the present experiment, since repumping only occurs in a
narrow frequency range (Fig.~\ref{fig2}(d)), when the excitation hits the resonance of NV$^{0}$.\\
\begin{figure*}[]
\centerline
{\includegraphics[]{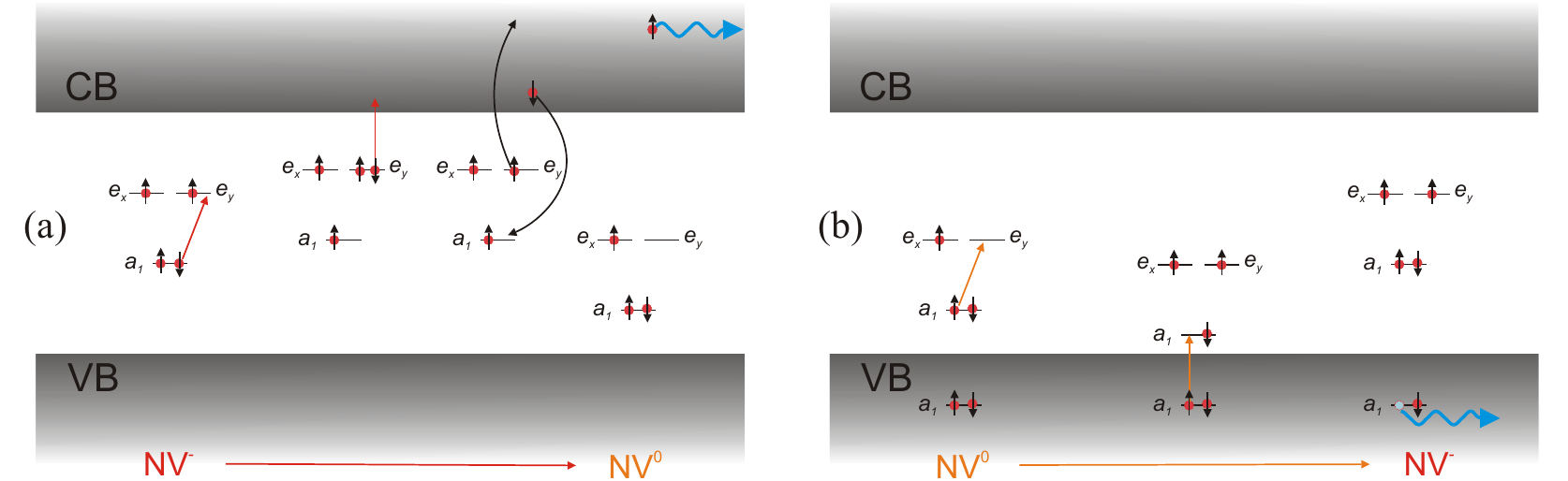}}
\caption{\label{fig4} Schematic picture of the charge conversion process for negative and neutral states of the NV defect. (a) NV$^{-}$ to NV$^{0}$ conversion involves two photons and an Auger process that releases enough energy to detach an electron from the defect. (b) NV$^{-}$ recovering occurs in two steps: first, an electron is excited from the {\it a$_{1}$} to the {\it e} orbital and then an additional electron is transferred form the deep lying {\it a$_{1}$} orbital to the vacant place on the {\it a$_{1}$} orbital in the band gap. The deep hole migrates away from the NV center leaving it in the negative charge state. The position of defect levels changes upon different occupation of states shown schematically for both processes.}
\end{figure*}
\indent Conventionally an off-resonant shorter wavelength laser (532~nm) is used to restore NV$^{-}$ emission after bleaching by resonant excitation. However, this results in an unwanted side effect -- spectral shift of the NV$^{-}$ transitions. High power (few milliwatts) 532~nm laser pulses can excite many closely situated impurities and thereby drastically change surrounding charge distribution. The spectral stability of NV$^{-}$ is improved when repumping is done at the ZPL of NV$^{0}$, since the less energetic and lower intensity yellow laser has lesser impact on the system. To quantify this effect, two scans were carried out with both types of repumping: green laser pulses between red laser sweeps, and a laser sweep over the NV$^{0}$ transition at the beginning of each red laser frequency sweep. Two typical scans for the same transition are shown in the Fig.~\ref{fig3}(a,b). A more than fourfold enhancement of the NV$^{-}$ spectral stability is achieved in the method exploiting repumping at ZPL of NV$^{0}$.\\
\indent Motivated by the new findings in our experiments we developed
a novel model based on hybrid density functional calculations.  We
propose that the photoionization of the NV$^{-}$ occurs in two steps
which involves a two photon absorption followed by an Auger process
(Fig.~\ref{fig4}(a)). The simultaneous absorption of two photons,
implied by the quadric dependence of the ionization rate with the
laser power, promotes an electron from the {\it a$_{1}$}
orbital in the band gap to the conduction band via the {\it e} level
in the gap. DFT calculations show that the minimum energy required for
such a transition is 3.5~eV, which is 0.4~eV smaller than the two photon
absorption energy of the excitation laser used at 637~nm. Relaxation from this high energy excited state to the ground state is
mediated by an Auger process and releases enough energy to excite an
electron $>$1.4~eV above the conduction band edge and detach it from the
defect. Hence, the NV defect remains in its neutral charge state. In order
to convert NV$^{0}$ back to its negative counterpart an electron needs to be
acquired. We assume that after exciting
an electron to the {\it e} orbital, a second excitation can occur
promoting an electron from the valence band into the vacated {\it
a$_{1}$} orbital in the gap. The analysis of the projected density
of state together with the calculated transition dipole moment shows a very
strong resonant transition from the {\it a$_{1}$} orbital in the
valence band to the {\it a$_{1}$} orbital in the gap, with an intensity comparable to that between the {\it a$_{1}$} and {\it e} orbitals in the gap. The position of the {\it a$_{1}$} orbital in the valence band indicates that the electron from this {\it a$_{1}$} orbital can be
promoted to the emptied {\it a$_{1}$} orbital in the gap by absorbing the second photon
with an energy at ZPL of NV$^{0}$. This finding may well explain the low power needed for the
recovery process. For NV$^{-}$ the first {\it $a_{1}$}
level lies deeper in the valence band making any excitation
from this level impossible with the range of energies considered here. Our model is supported by advanced DFT calculations and explains all the features of our experimental results.\\
\indent In conclusion, we have explicitly demonstrated that resonant
excitation neutralizes the NV$^{-}$ defect by direct observation of PLE
spectrum of its neutral counterpart. We have also shown that resonant
excitation of NV$^{0}$ assists in capturing an electron, thereby converting it into the negatively charged NV defect. This process occurs
even at very low power with little effect on the spectral line
position of the NV$^{-}$ center. This technique can be exploited instead of conventional
repumping with a frequency-doubled Nd:YAG laser in various low temperature experiments on NV centers, which require long-term stability of optical transitions~\cite{Togan_Nature2010,Robledo_Nature2011}. A new model of charged
state conversion was tentatively proposed, which does not involve any
additional impurities close to the NV defect and suggests that this is an
intrinsic process taking place on the defect itself. This study
provides a more comprehensive picture of the charge dynamics of the NV center
and explains for instance the nature of long living dark state
heretofore assigned to metastable state reported in
Ref.~\cite{Hell_NL2010} or prove an assumption in
Ref.~\cite{Waldherr_PRL2011} that the observed dark state is the neutral
NV defect.
\begin{acknowledgments}
We thank Ch. Santori and A. Batalov for motivating
discussion. Support by the Deutsche Forschungsgesellschaft (FOR 1482, FOR 1493, SFB/TR 21) and the European Commission EU FP7 Grant No.\ 270197 is gratefully acknowledged.
\end{acknowledgments}

\providecommand{\noopsort}[1]{}\providecommand{\singleletter}[1]{#1}%
\end{document}